\documentclass[12pt]{article}
\usepackage{amssymb}
\usepackage{epsf}
\usepackage{epsfig}
\newcommand{\lsim}{
\mathrel{\hbox{\rlap{\hbox{\lower4pt\hbox{$\sim$}}}\hbox{$<$}}}}

\newcommand{\gsim}{
\mathrel{\hbox{\rlap{\hbox{\lower4pt\hbox{$\sim$}}}\hbox{$>$}}}}

\newcommand{\be}{\begin{equation}}
\newcommand{\ee}{\end{equation}}
\newcommand{\bi}{\begin{itemize}}
\newcommand{\ei}{\end{itemize}}

\begin{document}
\begin{titlepage}
\vspace*{-0.5truecm}

\begin{flushright}
CERN-PH-TH/2004-146\\
TUM-HEP-556/04\\
hep-ph/0408016
\end{flushright}

\vspace*{2.0truecm}

\begin{center}
\boldmath
{\Large{\bf Waiting for the Discovery of $B^0_d\to K^0\bar K^0$}}
\end{center}

\vspace{0.7truecm}

\begin{center}
{\bf Robert Fleischer${}^a$  and  Stefan Recksiegel${}^b$}
 
\vspace{0.4truecm}

${}^a$ {\sl Theory Division, Department of Physics, CERN, 
CH-1211 Geneva 23, Switzerland}

\vspace{0.2truecm}

${}^b$ {\sl Physik Department, Technische Universit\"at M\"unchen,
D-85748 Garching, Germany}

\end{center}

\vspace{0.9cm}
\begin{abstract}
\vspace{0.2cm}\noindent
The CP asymmetries of the decay $B_d^0\to K^0\bar K^0$, which originates from 
$\bar b\to \bar d s \bar s$ flavour-changing neutral-current processes, 
and its CP-averaged branching ratio BR$(B_d\to K^0\bar K^0)$ offer 
interesting avenues to explore flavour physics. We show that we may 
characterize this channel, within the Standard Model, in a theoretically 
clean manner through a surface in observable space. In order to extract the 
relevant information from BR$(B_d\to K^0\bar K^0)$, further information is 
required, which is provided by the $B\to\pi\pi$ system and the $SU(3)$ 
flavour symmetry, where we include the leading factorizable $SU(3)$-breaking 
corrections and discuss how experimental insights into non-factorizable 
effects can be obtained. We point out that the Standard Model implies a 
{\it lower} bound for BR$(B_d\to K^0\bar K^0)$, which is very close to its 
current experimental upper bound, thereby suggesting that this decay should 
soon be observed. Moreover, we explore the implications for ``colour 
suppression'' in the $B\to\pi\pi$ system, and convert the data for these 
modes into a peculiar Standard-Model pattern for the CP-violating 
$B_d^0\to K^0\bar K^0$ observables.
\end{abstract}

\vspace*{0.5truecm}
\vfill
\noindent
August 2004

\end{titlepage}

\thispagestyle{empty}
\vbox{}
\newpage

\setcounter{page}{1}

\section{Setting the Stage}
%
%
The $B$ factories allow us to confront the Kobayashi--Maskawa (KM) mechanism 
of CP violation \cite{KM}, which describes this phenomenon in the Standard 
Model (SM), with a steadily increasing amount of experimental data (for a 
recent overview, see \cite{RF-Armenia}). An interesting element of this 
programme is the decay $B_d^0\to K^0\bar K^0$. It originates from
$\bar b\to \bar d s \bar s$ flavour-changing neutral-current (FCNC)
processes, which are governed by QCD penguin diagrams in the SM.
Should these topologies be dominated by internal top-quark exchanges, 
the CP asymmetries of $B_d^0\to K^0\bar K^0$ would vanish in the SM thanks
to a subtle cancellation of weak phases, thereby suggesting an interesting 
test of the KM mechanism (see, for instance, \cite{quinn}). However, 
contributions from penguins with internal up- and charm-quark exchanges are 
expected to yield sizeable CP asymmetries in $B_d^0\to K^0\bar K^0$ even 
within the SM, so that the interpretation of these effects is much more 
complicated \cite{RF-BdKK}. In view of the impressive progress since these 
early studies of $B_d^0\to K^0\bar K^0$, and the strong experimental upper 
bound for the corresponding CP-averaged branching ratio \cite{HFAG},
\begin{equation}\label{BR-CP-av}
\mbox{BR}(B_d\to K^0\bar K^0)\equiv 
\frac{\mbox{BR}(B_d^0\to K^0\bar K^0)+
\mbox{BR}(\bar B_d^0\to K^0\bar K^0)}{2} < 1.5\times 10^{-6} \,\,
\mbox{(90\% C.L.)},
\end{equation}
it is interesting to return to this decay. 

As usual, we consider the following time-dependent rate asymmetry:
\begin{eqnarray}
\lefteqn{\frac{\Gamma(B^0_d(t)\to  K^0\bar K^0)-\Gamma(\bar B^0_d(t)\to 
K^0\bar K^0)}{\Gamma(B^0_d(t)\to  K^0\bar K^0)+\Gamma(\bar B^0_d(t)\to 
K^0\bar K^0)}}\nonumber\\
&&={\cal A}_{\rm CP}^{\rm dir}(B_d\to  K^0\bar K^0)\cos(\Delta M_d t)+
{\cal A}_{\rm CP}^{\rm mix}(B_d\to  K^0\bar K^0)
\sin(\Delta M_d t),\label{BdKK-rate-asym}
\end{eqnarray}
where ${\cal A}_{\rm CP}^{\rm dir}(B_d\to  K^0\bar K^0)$ and
${\cal A}_{\rm CP}^{\rm mix}(B_d\to  K^0\bar K^0)$ describe the
direct and mixing-induced CP asymmetries, respectively. In 
order to analyse these observables, we have to parametrize the
$B_d^0\to K^0\bar K^0$ decay amplitude appropriately. Within the
SM, we may write
\begin{equation}\label{ampl-BdKK}
A(B_d^0\to K^0\bar K^0)=\lambda^{(d)}_u {\cal P}_u^{K\!K} + 
\lambda^{(d)}_c {\cal P}_c^{K\!K} +  \lambda^{(d)}_t {\cal P}_t^{K\!K},
\end{equation}
where the $\lambda^{(d)}_q \equiv V_{qd}V_{qb}^\ast$ are CKM factors,
and the ${\cal P}_q^{K\!K}$ denote the strong amplitudes of penguin topologies
with internal $q$-quark exchanges, which receive tiny contributions from
colour-suppressed electroweak (EW) penguins and are fully dominated by 
QCD penguin processes. If we now eliminate $\lambda^{(d)}_t$ with the
help of the relation
\begin{equation}
\lambda^{(d)}_t=-\lambda^{(d)}_u-\lambda^{(d)}_c,
\end{equation}
which follows from the unitarity of the Cabibbo--Kobayashi--Maskawa
(CKM) matrix, and use the Wolfenstein parametrization \cite{wolf}, 
we obtain
\begin{equation}\label{ampl-BdKK-lamt}
A(B^0_d\to K^0\bar K^0)=\lambda^3A{\cal P}_{tc}^{K\!K}
\left[1-\rho_{K\!K} e^{i\theta_{K\!K}}e^{i\gamma}\right],
\end{equation}
where ${\cal P}_{tc}^{K\!K}\equiv {\cal P}_t^{K\!K}-{\cal P}_c^{K\!K}$, and 
\begin{equation}\label{rho-KK-def}
\rho_{K\!K} e^{i\theta_{K\!K}}\equiv R_b
\left[\frac{{\cal P}_t^{K\!K}-{\cal P}_u^{K\!K}}{{\cal P}_t^{K\!K}-
{\cal P}_c^{K\!K}}\right],
\end{equation}
with
\begin{equation}
R_b\equiv\left(1-\frac{\lambda^2}{2}\right)\frac{1}{\lambda}
\left|\frac{V_{ub}}{V_{cb}}\right|=\sqrt{\bar\varrho^2+\bar\eta^2}=0.37\pm0.04.
\end{equation}
Applying the standard formalism to deal with the CP-violating observables
provided by (\ref{BdKK-rate-asym}) \cite{RF-Armenia}, we straightforwardly 
arrive at
\begin{equation}\label{Adir-BdK0K0}
{\cal A}_{\rm CP}^{\rm dir}\equiv
{\cal A}_{\rm CP}^{\rm dir}(B_d\to K^0\bar K^0)=
\frac{2\rho_{K\!K}\sin\theta_{K\!K}\sin\gamma}{1-2\rho_{K\!K}\cos\theta_{K\!K}
\cos\gamma+\rho_{K\!K}^2}
\end{equation}
\begin{equation}\label{Amix-BdK0K0}
{\cal A}_{\rm CP}^{\rm mix}\equiv
{\cal A}_{\rm CP}^{\rm mix}(B_d\to K^0\bar K^0)=
\frac{\sin\phi_d-2\rho_{K\!K}\cos\theta_{K\!K}\sin(\phi_d+\gamma)
+\rho_{K\!K}^2\sin(\phi_d+2\gamma)}{1-2\rho_{K\!K}\cos\theta_{K\!K}\cos\gamma+
\rho_{K\!K}^2},
\end{equation}
where the $B^0_d$--$\bar B^0_d$ mixing phase $\phi_d$ agrees with 
$2\beta$ in the SM; $\beta$ and $\gamma$ are the usual angles of
the unitarity triangle of the CKM matrix. 

The outline of this paper is as follows: in Section~\ref{sec:char-SM},
we show that $B^0_d\to K^0\bar K^0$ can be efficiently characterized in 
the SM through a surface in the three-dimensional space of its observables. 
In order to extract the relevant information from the CP-averaged branching 
ratio, an additional input is needed, which is offered by the $B\to\pi\pi$ 
system and the $SU(3)$ flavour symmetry. We show how insights into 
non-factorizable $SU(3)$-breaking effects in the relevant hadronic penguin 
amplitudes can be obtained, and point out that the current $B$-factory data 
are consistent with small corrections, although the experimental 
uncertainties are still large. One of the main results of our analysis are 
{\it lower} bounds for BR$(B_d\to K^0\bar K^0)$, which are remarkably close
to the experimental {\it upper} bound in (\ref{BR-CP-av}), thereby 
suggesting that this decay should be observed in the near future at the 
$B$ factories. In Section~\ref{sec:Bpipi-BKK}, we demonstrate then that 
the measurement of the $B_d\to K^0\bar K^0$ observables will allow us to 
reveal the hadronic substructure of the $B\to\pi\pi$ system, providing in 
particular insights into the issue of ``colour suppression''. Conversely,
using the pattern of the current $B$-factory data as a guideline, we 
calculate allowed regions in the space of the CP-violating 
$B_d\to K^0\bar K^0$ observables within the SM, which may be helpful in 
the future to search for new-physics (NP) contributions to 
$\bar b\to \bar d s \bar s$ FCNC processes. Finally, we summarize our 
conclusions in Section~\ref{sec:concl}.

\boldmath
\section{Standard-Model Picture of 
$B^0_d\to K^0\bar K^0$}\label{sec:char-SM}
\unboldmath
\subsection{Preliminaries: Top-Quark Dominance}
It is instructive to have first a brief look at the case of top-quark
dominance, where (\ref{rho-KK-def}) simplifies as follows:
\begin{equation}\label{rhoKK-top}
\rho_{K\!K} e^{i\theta_{K\!K}}=R_b.
\end{equation}
Since the CP-conserving strong phase $\theta_{K\!K}$ vanishes in this
expression, (\ref{Adir-BdK0K0}) implies that the direct CP asymmetry 
of $B_d\to K^0\bar K^0$ vanishes as well.
The analysis of the mixing-induced CP asymmetry (\ref{Amix-BdK0K0})
is a bit more complicated. If we take into account that we
have $\phi_d=2\beta$ in the SM, and use the relations
\begin{equation}
\sin\beta=\frac{\bar\eta}{\sqrt{(1-\bar\varrho)^2+\bar\eta^2}}, \quad
\cos\beta=\frac{1-\bar\varrho}{\sqrt{(1-\bar\varrho)^2+\bar\eta^2}}
\end{equation}
\begin{equation}
\sin\gamma=\frac{\bar\eta}{\sqrt{\bar\varrho^2+\bar\eta^2}}, \quad
\cos\gamma=\frac{\bar\varrho}{\sqrt{\bar\varrho^2+\bar\eta^2}}
\end{equation}
between the angles of the unitarity triangle and the Wolfenstein parameters
\cite{wolf}, we may show that ${\cal A}_{\rm CP}^{\rm mix}$ would actually 
also vanish. This can be seen more transparently if we eliminate 
$\lambda^{(d)}_u$ instead of $\lambda^{(d)}_t$ in (\ref{ampl-BdKK}). 
Assuming then top-quark dominance, we obtain a cancellation between the 
weak phase $\beta$ of $\lambda^{(d)}_t$ and the $\beta$ introduced through 
the SM value of $\phi_d$, thereby yielding a vanishing mixing-induced 
$B^0_d\to K^0\bar K^0$ CP asymmetry~\cite{quinn}. For our purposes, the 
parametrization in (\ref{ampl-BdKK-lamt}) is, however, more appropriate.

\begin{figure}
\vspace*{0.3truecm}
\begin{center}
\includegraphics[width=15.8cm]{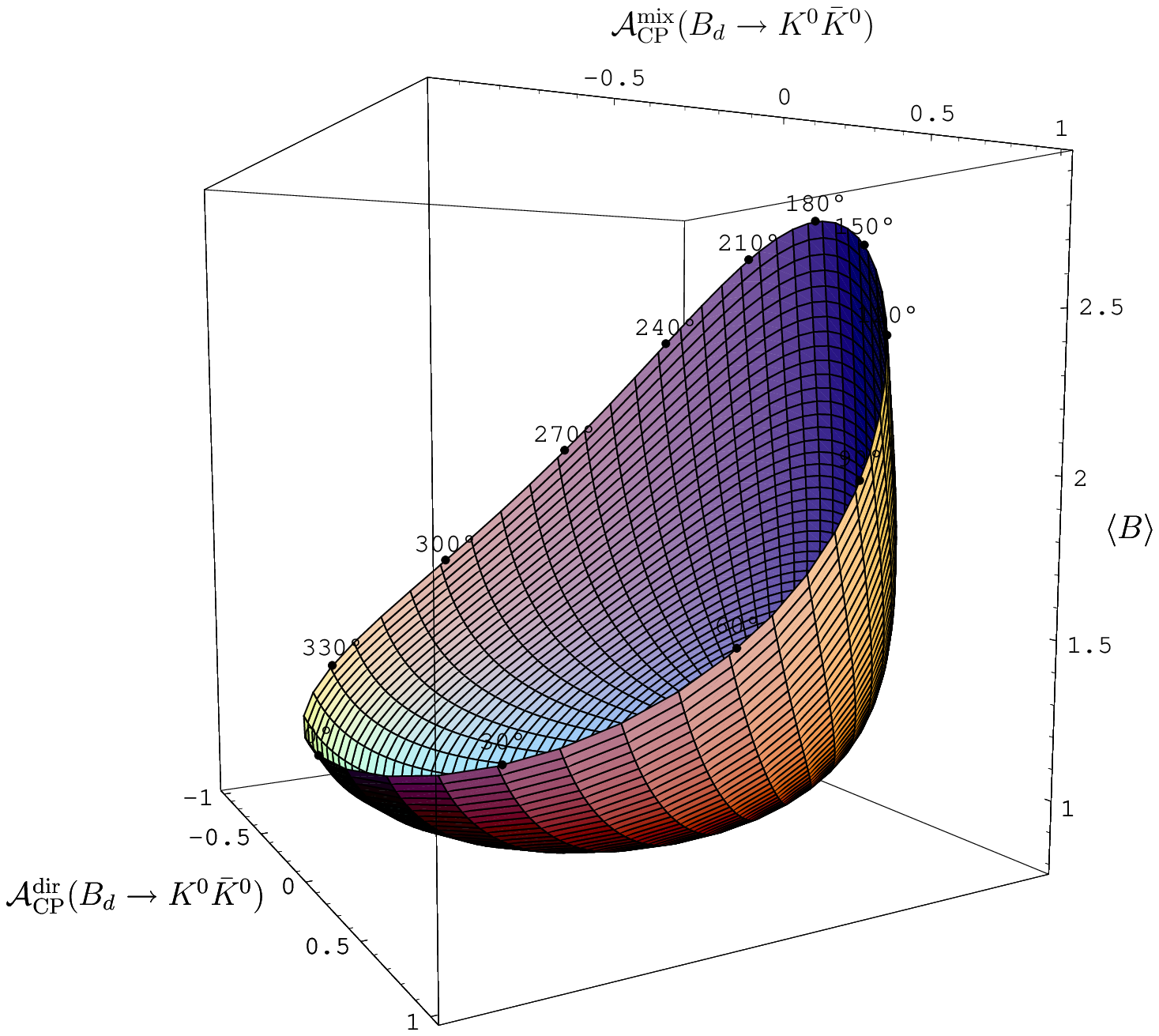}
\end{center}
\caption{The surface in the 
${\cal A}_{\rm CP}^{\rm dir}$--${\cal A}_{\rm CP}^{\rm 
mix}$--$\langle B \rangle$ observable space of 
$B^0_d\to K^0\bar K^0$ for $\phi_d = 47^\circ$ and $\gamma=65^\circ$, 
characterizing this decay in the SM. The intersecting lines on the surface 
correspond to constant $\rho_{K\!K}$ and $\theta_{K\!K}$, respectively. 
The numbers on the fringe indicate the value of $\theta_{K\!K}$, the fringe 
itself is defined by $\rho_{K\!K}=1$. \label{fig:SM-surface}}
\end{figure}

\subsection{Characteristic Surface in Observable Space}
\begin{figure}
\vspace*{0.3truecm}
\begin{center}
\includegraphics[width=6.7cm]{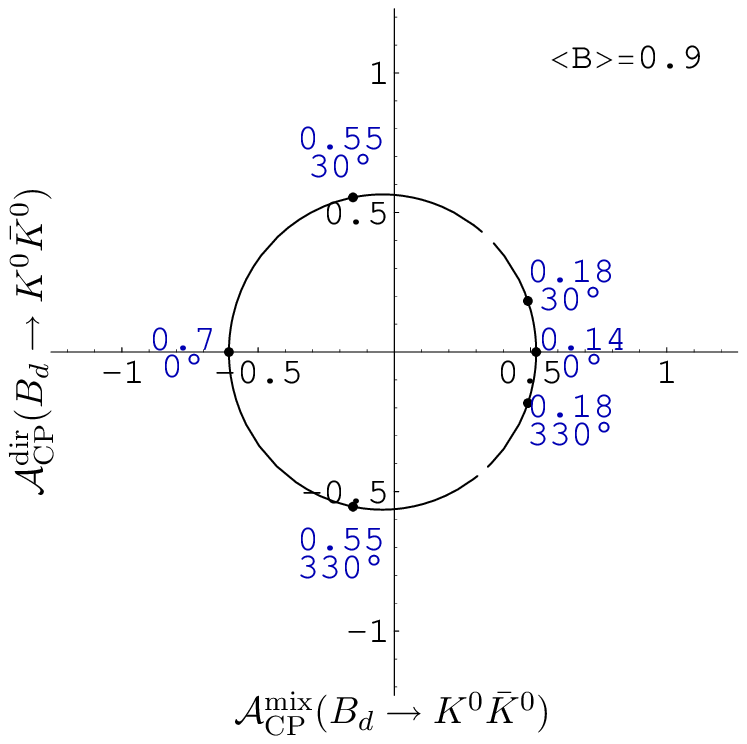}
\includegraphics[width=6.7cm]{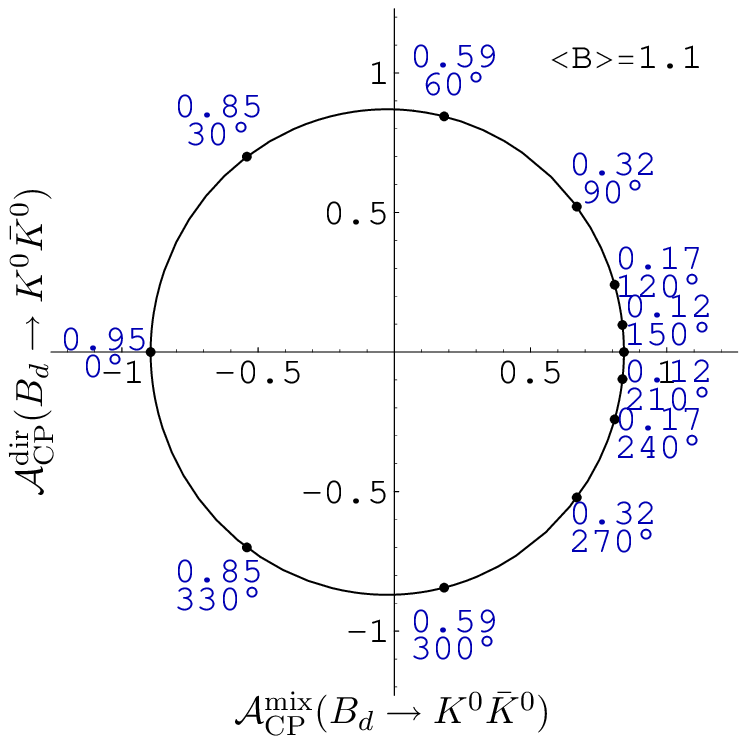}
\includegraphics[width=6.7cm]{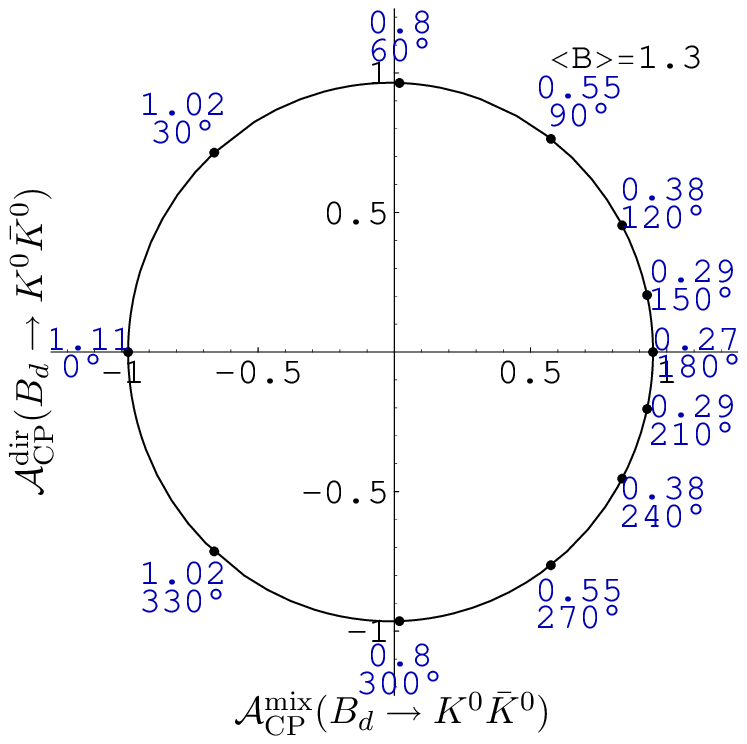}
\end{center}
\caption{The ellipses arising in the 
${\cal A}_{\rm CP}^{\rm mix}$--${\cal A}_{\rm CP}^{\rm dir}$ plane
for given values of $\langle B \rangle$, with the associated 
values of $\rho_{K\!K}$ and $\theta_{K\!K}$. As in Fig.~\ref{fig:SM-surface},
we have chosen $\phi_d = 47^\circ$ and $\gamma=65^\circ$.}\label{fig:ellipses}
\end{figure}

In the following analysis, we assume that
\begin{equation}\label{SM-CKM-phases}
\phi_d=2\beta=(47\pm4)^\circ, \quad \gamma=(65\pm7)^\circ,
\end{equation}
as in the SM \cite{CKM-Book}. By the time the CP-violating asymmetries 
in (\ref{Adir-BdK0K0}) and (\ref{Amix-BdK0K0}) can be reliably measured, 
the picture of these parameters will be much sharper. The measurement of 
${\cal A}_{\rm CP}^{\rm dir}$ and ${\cal A}_{\rm CP}^{\rm mix}$ allows us 
then to extract the hadronic parameters $\rho_{K\!K}$ and $\theta_{K\!K}$ 
in a {\it theoretically clean} manner. Although these quantities are
interesting for the analysis of charged $B\to\pi K$ modes, as we will
see below, and can nicely be compared with theoretical predictions, such 
as those of the ``QCD factorization'' approach \cite{QCDF}, they do not 
provide -- by themselves -- a test of the SM description of the 
$\bar b\to\bar d s \bar s$ FCNC processes mediating the decay 
$B^0_d\to K^0\bar K^0$. However, so far, 
we have not yet used the information offered by the CP-averaged branching 
ratio introduced in (\ref{BR-CP-av}). The parametrization in 
(\ref{ampl-BdKK-lamt}) allows us to write
\begin{equation}\label{BR-BKK-expr}
\mbox{BR}(B_d\to K^0\bar K^0)=\frac{\tau_{B_d}}{16\pi M_{B_d}}
\Phi(M_{K}/M_{B_d},M_{K}/M_{B_d}) 
|\lambda^3 A {\cal P}_{tc}^{K\!K}|^2 \langle B \rangle,
\end{equation}
where 
\begin{equation}
\Phi(x,y)=\sqrt{\left[1-(x+y)^2\right]\left[1-(x-y)^2\right]}
\end{equation}
is the two-body phase-space function, and
\begin{equation}\label{B-DEF}
\langle B \rangle\equiv 1-2\rho_{K\!K}\cos\theta_{K\!K}
\cos\gamma+\rho_{K\!K}^2.
\end{equation}
If we now use the SM values of $\phi_d$ and $\gamma$, we may characterize
the decay $B^0_d\to K^0\bar K^0$ -- within the SM -- through a surface in 
the observable space of ${\cal A}_{\rm CP}^{\rm dir}$, 
${\cal A}_{\rm CP}^{\rm mix}$ and $\langle B \rangle$. In 
Fig.~\ref{fig:SM-surface}, we show this surface, where each point 
corresponds to a given value of $\rho_{K\!K}$ and $\theta_{K\!K}$. It should 
be emphasized that this surface is {\it theoretically clean} since it 
relies only on the general SM parametrization of $B^0_d\to K^0\bar K^0$. 
Consequently, should future measurements give a value in observable space 
that should {\it not} lie on the SM surface, we would have immediate evidence 
for NP contributions to $\bar b\to \bar d s \bar s$ FCNC processes. If
we consider a fixed value of $\langle B \rangle$, we obtain ellipses
in the ${\cal A}_{\rm CP}^{\rm dir}$--${\cal A}_{\rm CP}^{\rm mix}$
plane, which are described by
\begin{equation}\label{B-ellipses}
\left[\frac{{\cal A}_{\rm CP}^{\rm dir}}{a_{{\cal A}_{\rm CP}^{\rm dir}}}
\right]^2+\left[\frac{{\cal A}_{\rm CP}^{\rm mix}-
{\cal A}_0}{a_{{\cal A}_{\rm CP}^{\rm mix}}}\right]^2=1,
\end{equation}
with
\begin{equation}
{\cal A}_0=\left[
\frac{\langle B \rangle -2 \sin^2\gamma}{\langle B \rangle}
\right]\sin(\phi_d+2\gamma)
\end{equation}
and 
\begin{equation}\label{ellipses2}
a_{{\cal A}_{\rm CP}^{\rm dir}}=2
\frac{\sqrt{\langle B\rangle -\sin^2\gamma}}{\langle B\rangle}
|\sin\gamma|,
\quad a_{{\cal A}_{\rm CP}^{\rm mix}} = a_{{\cal A}_{\rm CP}^{\rm dir}}
|\cos(\phi_d+2\gamma)|.
\end{equation}
In Fig.~\ref{fig:ellipses}, we show these ellipses for various values
of $\langle B\rangle$. Since $\sin(\phi_d+2\gamma)=0.05$ and 
$\cos(\phi_d+2\gamma)=-1.00$ for the central values of (\ref{SM-CKM-phases}), 
we have actually to deal -- to a good approximation -- with circles 
around the origin in the case of this figure.

In the derivation of (\ref{B-ellipses}), we have assumed that
$\langle B\rangle -\sin^2\gamma>0$, which enters in (\ref{ellipses2}).
In fact, if we consider (\ref{B-DEF}) and vary $\rho_{K\!K}$ and 
$\theta_{K\!K}$ as free parameters, while keeping $\gamma$ fixed, we find 
that $\langle B \rangle$ takes the following {\it absolute} minimum:
\begin{equation}\label{B-min}
\langle B \rangle_{\rm min}=\sin^2\gamma=0.82\,^{+0.08}_{-0.10},
\end{equation}
which corresponds to
\begin{equation}
\rho_{K\!K}=\cos\gamma=0.42 \pm 0.11, 
\quad \theta_{K\!K}=0^\circ, 
\end{equation}
yielding
\begin{equation}\label{CP-obs-extr}
{\cal A}_{\rm CP}^{\rm dir}=0, \quad
{\cal A}_{\rm CP}^{\rm mix}=-\sin(\phi_d+2\gamma)=
-\left(0.05 \pm 0.25 \right).
\end{equation}
The numerical results in (\ref{B-min})--(\ref{CP-obs-extr}) were calculated
with the help of (\ref{SM-CKM-phases}). It is amusing to note that the 
associated values of $\rho_{K\!K}$ and $\theta_{K\!K}$ are very close to 
the case of top-quark dominance, as can be seen in (\ref{rhoKK-top}).

\boldmath
\subsection{Extraction of $\langle B \rangle$}
\unboldmath
Whereas ${\cal A}_{\rm CP}^{\rm dir}$ and ${\cal A}_{\rm CP}^{\rm mix}$
can be directly obtained from (\ref{BdKK-rate-asym}), the 
extraction of $\langle B\rangle$ from (\ref{BR-BKK-expr}) requires 
additional information. To this end, we follow \cite{BF-BdKK},
and combine $B_d\to K^0\bar K^0$ with $B_d\to\pi^+\pi^-$. It is then 
useful to write the decay amplitude of the latter mode as 
\begin{equation}\label{Bdpipi-ampl}
A(B^0_d\to\pi^+\pi^-)=-\lambda^3 A {\cal P}_{tc}^{\pi\pi}
\left[1-e^{i\gamma}\frac{1}{d e^{i\theta}}\right],
\end{equation}
where ${\cal P}_{tc}^{\pi\pi}$ is the $B_d\to\pi^+\pi^-$
counterpart of ${\cal P}_{tc}^{K\!K}$, and $d e^{i\theta}$ is
a hadronic parameter. Performing an isospin analysis of the
$B\to\pi\pi$ system for the SM values of $\phi_d$ and $\gamma$ in 
(\ref{SM-CKM-phases}), $d$ and $\theta$ could be extracted from the 
$B$-factory data, with the following result \cite{BFRS}:
\begin{equation}\label{d-theta-extr}
d=0.48\,^{+0.35}_{-0.22}, \quad \theta=+\left(138\,^{+19}_{-23}\right)^\circ;
\end{equation}
similar values were subsequently obtained in \cite{Bpipi-alt}. If we 
calculate now the CP-averaged $B_d\to\pi^+\pi^-$ branching ratio
with the help of (\ref{Bdpipi-ampl}), (\ref{BR-BKK-expr}) implies
\begin{equation}\label{B-extr-1}
\langle B \rangle=
\left|\frac{{\cal P}_{tc}^{\pi\pi}}{{\cal P}_{tc}^{K\!K}}\right|^2
\left[\frac{\mbox{BR}(B_d\to K^0\bar K^0)}{\mbox{BR}(B_d\to\pi^+\pi^-)}
\right]F_{\pi\pi}(d,\theta),
\end{equation}
where we have introduced
\begin{equation}\label{F-def}
F_{\pi\pi}(d,\theta)\equiv \frac{1-2d\cos\theta\cos\gamma+d^2}{d^2}=
6.57\,^{+6.65}_{-4.20},
\end{equation}
and have neglected tiny phase-space differences. The numerical value
in (\ref{F-def}) follows from the $B\to\pi\pi$ analysis performed in 
\cite{BFRS}. 
In the future, the corresponding uncertainties, which are only of
experimental origin, can be reduced considerably. Let us emphasize
that (\ref{B-extr-1}) is valid {\it exactly} in the SM. In order
to deal with the $|{\cal P}_{tc}^{\pi\pi}/ {\cal P}_{tc}^{K\!K}|$ 
factor, we neglect colour-suppressed EW penguins, which have an
essentially negligible impact on the $B_d\to K^0\bar K^0$ and
$B_d\to\pi^+\pi^-$ modes \cite{GHLR-EWP}, and use the $SU(3)$ flavour 
symmetry of strong interactions. In the strict $SU(3)$ limit, this ratio 
equals one. If we take the factorizable $SU(3)$-breaking corrections into 
account,\footnote{Chiral terms can be related through the Gell-Mann--Okubo 
relation, as discussed in \cite{RF-BsKK}.} we obtain
\begin{equation}\label{SU(3)-fact1}
\left|\frac{{\cal P}_{tc}^{\pi\pi}}{{\cal P}_{tc}^{K\!K}}\right|_{\rm fact}
=\left[\frac{f_\pi F_{B\pi}(M_\pi^2;0^+)}{f_K F_{B K}(M_K^2;0^+)}\right]
\left[\frac{M_B^2-M_\pi^2}{M_B^2-M_K^2}\right]=0.64,
\end{equation}
where $f_\pi=131\,{\rm MeV}$ and $f_K=160\,{\rm MeV}$ denote the pion and 
kaon decay constants, and the form factors $F_{B\pi}(M_\pi^2;0^+)$ and
$F_{B K}(M_K^2;0^+)$ parametrize the hadronic quark-current matrix
elements $\langle \pi^-|(\bar b u)_{\rm V-A}| B^0_d\rangle$ 
and $\langle K^0|(\bar b s)_{\rm V-A}| B^0_d\rangle$, respectively.
The numerical value in (\ref{SU(3)-fact1}) corresponds to the light-cone
sum-rule analysis performed recently in \cite{Ball} (with $\delta_{a1}=0$), 
while the form factors obtained within the Bauer--Stech--Wirbel (BSW)
model \cite{BSW} yield a value of 0.72.

\boldmath
\subsection{Exploring Non-Factorizable $SU(3)$-Breaking Corrections}
\unboldmath
Insights into the issue of factorization and $SU(3)$-breaking effects 
of the hadronic ${\cal P}_{tc}$ penguin amplitudes can be obtained with 
the help of $B\to\pi K$ modes, which originate from $\bar b\to\bar s$ 
quark-level processes. Applying the formalism of \cite{BFRS}, we write
\begin{equation}\label{BpiK-Bpipi}
\left[\frac{\mbox{BR}(B^\pm\to\pi^\pm K)}{\mbox{BR}(B_d\to\pi^+\pi^-)}\right]
\left[\frac{\tau_{B_d}}{\tau_{B^+}}\right]=\frac{1}{\epsilon}
\left|\frac{{\cal P}_{tc}^{\pi K}}{{\cal P}_{tc}^{\pi\pi}}\right|^2
\left[\frac{1+\delta R}{F_{\pi\pi}(d,\theta)}\right],
\end{equation}
where 
\begin{equation}
\epsilon\equiv\frac{\lambda^2}{1-\lambda^2}=0.05,
\end{equation}
and
\begin{equation}
\delta R = 2\rho_{\rm c}\cos\theta_{\rm c}\cos\gamma+\rho_{\rm c}^2
-2\left[\cos\psi_{\rm C}^{(1)} + \rho_{\rm c}
\cos(\theta_{\rm c}-\psi_{\rm C}^{(1)})\cos\gamma\right] 
a_{\rm EW}^{{\rm C}(1)} + \left[a_{\rm EW}^{{\rm C}(1)}\right]^2.
\end{equation}
The hadronic parameter $\rho_{\rm c}e^{i\theta_{\rm c}}$ is the
$B^+\to\pi^+K^0$ counterpart of $\rho_{K\!K} e^{i\theta_{K\!K}}$. Because of 
the different CKM structure of $B^+\to\pi^+K^0$, we have
\begin{equation}\label{rho-c}
\rho_{\rm c}e^{i\theta_{\rm c}}\approx 
\epsilon\,\rho_{K\!K} e^{i\theta_{K\!K}},
\end{equation}
so that $\rho_{\rm c}$ is expected at the few percent level. 
The parameter $a_{\rm EW}^{{\rm C}(1)}$ and the strong phase 
$\psi_{\rm C}^{(1)}$ are related to colour-suppressed EW penguins. 
It is expected that $a_{\rm EW}^{{\rm C}(1)}$ is also of ${\cal O}(10^{-2})$. 
Interestingly, the analysis performed in \cite{BFRS} allows us to 
determine $\delta R$ from the data with the help of the following 
relation:
\begin{equation}\label{1pdelR}
1+\delta R=\frac{1-2r\cos\delta\cos\gamma+r^2}{R},
\end{equation}
where
\begin{equation}
R\equiv\left[\frac{\mbox{BR}(B_d^0\to\pi^- K^+)+
\mbox{BR}(\bar B_d^0\to\pi^+ K^-)}{\mbox{BR}(B^+\to\pi^+ K^0)+
\mbox{BR}(B^-\to\pi^- \bar K^0)}
\right]\frac{\tau_{B^+}}{\tau_{B_d}}
=0.91\pm0.07,
\end{equation}
and the hadronic parameters
\begin{equation}\label{r-det-Bpipi}
r=0.11^{+0.07}_{-0.05},\quad \delta=+(42^{+23}_{-19})^\circ
\end{equation}
were fixed through
\begin{equation}\label{r-d-rel}
re^{i\delta}=\frac{\epsilon}{d}e^{i(\pi-\theta)}
\end{equation}
from the $B\to\pi\pi$ analysis, which yields (\ref{d-theta-extr}).
Following these lines, we obtain
\begin{equation}\label{deltaR}
\delta R= 0.036 \,^{+0.094}_{-0.079},
\end{equation}
which is nicely complemented by the experimental results \cite{HFAG} 
for the direct CP asymmetry
\begin{equation}
{\cal A}_{\rm CP}^{\rm dir}(B^\pm\to\pi^\pm K)\equiv
\frac{\mbox{BR}(B^+\to\pi^+K^0)-
\mbox{BR}(B^-\to\pi^-\bar K^0)}{\mbox{BR}(B^+\to\pi^+K^0)+
\mbox{BR}(B^-\to\pi^-\bar K^0)}=-0.02\pm0.06,
\end{equation}
taking the following form:
\begin{equation}\label{ACP-B+pi+K0}
{\cal A}_{\rm CP}^{\rm dir}(B^\pm\to\pi^\pm K)=
-2\rho_{\rm c}\left[\frac{\sin\theta_{\rm c}-
a_{\rm EW}^{{\rm C}(1)}\sin(\theta_{\rm c}-
\psi_{\rm C}^{(1)})}{1+\delta R}\right]\sin\gamma.
\end{equation}
Consequently, we have {\it no} experimental evidence for anomalously
large values of $\rho_{\rm c}$ and $a_{\rm EW}^{{\rm C}(1)}$. 
In particular, we do not find indications for an enhancement of 
the latter parameter describing the colour-suppressed EW penguin 
contributions, in contrast to the claims made recently in 
\cite{CKMfitters}.

If we write now
\begin{equation}
\left|\frac{{\cal P}_{tc}^{\pi\pi}}{{\cal P}_{tc}^{\pi K}}\right|
=\xi_{SU(3)}^{\rm n-fact}
\left|\frac{{\cal P}_{tc}^{\pi\pi}}{{\cal P}_{tc}^{\pi K}}\right|_{\rm fact}
\quad\mbox{with}\quad
\left|\frac{{\cal P}_{tc}^{\pi\pi}}{{\cal P}_{tc}^{\pi K}}\right|_{\rm fact}
=\frac{f_\pi}{f_K},
\end{equation}
we obtain from (\ref{BpiK-Bpipi}) with the help of (\ref{1pdelR}) 
and (\ref{r-d-rel})
\begin{equation}\label{xi-SU3-det}
\xi_{SU(3)}^{\rm n-fact}=
\frac{f_K}{f_\pi}\sqrt{\frac{1}{\epsilon}
\left[\frac{d^2+2\epsilon d\cos\theta\cos\gamma+\epsilon^2}{1-
2d\cos\theta\cos\gamma+d^2}\right]
\left[\frac{\mbox{BR}(B_d\to\pi^+\pi^-)}{\mbox{BR}(B_d\to\pi^\mp K^\pm)}
\right]}=1.01^{+0.48}_{-0.37},
\end{equation}
where the numerical value follows from the analysis in \cite{BFRS}. 
The current $B$-factory data do therefore not indicate a deviation
of $\xi_{SU(3)}^{\rm n-fact}$ from one, although the uncertainties
are still large. In the future, (\ref{xi-SU3-det}) can be determined
with much better accuracy. In particular, since this expression involves 
only $B$ decays with charged pions and kaons in the final state,\footnote{The
determination of $d$ and $\theta$ relies only on the measurement of the 
CP-violating $B_d\to\pi^+\pi^-$ observables, yielding a twofold solution.
Using additional information on the CP-averaged $B_d\to\pi^0\pi^0$ 
branching ratio, this ambiguity can be resolved, thereby yielding
the single solution in (\ref{d-theta-extr}) \cite{BFRS}.} 
it should be possible to explore it in a powerful way at LHCb 
\cite{LHC-Book}. A similar comment applies to the determination
of (\ref{F-def}). It should be noted that (\ref{xi-SU3-det}) does 
actually not only probe non-factorizable $SU(3)$-breaking effects, 
but also the importance of penguin annihilation topologies, which
contribute to $B_d\to\pi^+\pi^-$ and $B_d\to K^0\bar K^0$ (and are 
implicitly included in ${\cal P}_{tc}^{\pi\pi}$ and 
${\cal P}_{tc}^{K\!K}$, respectively), but do {\it not} contribute to 
${\cal P}_{tc}^{\pi K}$. Their importance can be explored through
the $B_d\to K^+K^-$, $B_s\to \pi^+\pi^-$ system. The experimental
upper bounds on the former decay \cite{BFRS}, as well as the numerical
value in (\ref{xi-SU3-det}), do not indicate any enhancement.

\boldmath
\subsection{Lower Bounds on the $B_d\to K^0\bar K^0$ Branching Ratio}
\unboldmath
By the time all $B_d\to K^0 \bar K^0$ observables can be measured with
a reasonable accuracy, we will have a good picture of (\ref{xi-SU3-det}).
We may then extrapolate correspondingly to the determination of
$|{\cal P}_{tc}^{\pi\pi}/ {\cal P}_{tc}^{K\!K}|$ through (\ref{SU(3)-fact1}),
allowing us to relate $\langle B\rangle$ to the CP-averaged
$B_d\to K^0\bar K^0$ branching ratio with the help of (\ref{B-extr-1}). For 
the following analysis, we will just use (\ref{SU(3)-fact1}), complementing
it with the numerical result in (\ref{F-def}) and 
$\mbox{BR}(B_d\to\pi^+\pi^-)=(4.6\pm0.4)\times10^{-6}$ \cite{HFAG}. We
are then in a position to convert the lower bound in (\ref{B-min}) into
the following lower bound for the CP-averaged $B_d\to K^0\bar K^0$ 
branching ratio:
\begin{equation}\label{BRmin1}
\mbox{BR}(B_d\to K^0\bar K^0)_{\rm min} = \left(1.39^{+1.54}_{-0.95}\right)   
\times\left[\frac{F_{B K}(M_K^2;0^+)}{0.331}\frac{0.258}{F_{B\pi}
(M_\pi^2;0^+)}\right]^2\times 10^{-6}.
\end{equation}
In this expression, we made the dependence on the form factors
explicit, where the numerical values refer to \cite{Ball}. If
we use the BSW form factors \cite{BSW}, the lower bound on 
$\mbox{BR}(B_d\to K^0\bar K^0)$ is reduced by about $20\%$.

Interestingly, a picture similar to the one of (\ref{BRmin1}) 
emerges also from a very different avenue: it is a nice 
feature of (\ref{B-extr-1}) that this relation uses only 
$\bar b\to \bar d$ transitions. However, it is also useful to combine
$B^0_d\to K^0\bar K^0$ with the $\bar b\to \bar s$ transition
$B^+\to\pi^+K^0$. As we have noted above, in doing this we have
to neglect the penguin annihilation topologies contributing to 
the former mode. Neglecting phase-space differences for simplicity, 
we may then write
\begin{equation}\label{B-det2}
\langle B \rangle = \frac{1}{\epsilon}
\left|\frac{{\cal P}_{tc}^{\pi K}}{{\cal P}_{tc}^{K\!K}}\right|^2
(1+\delta R)\left[
\frac{\mbox{BR}(B_d\to K^0\bar K^0)}{\mbox{BR}(B^\pm\to\pi^\pm K)}
\right]\frac{\tau_{B^+}}{\tau_{B_d}},
\end{equation}
where
\begin{equation}\label{SU(3)-fact2}
\left|\frac{{\cal P}_{tc}^{\pi K}}{{\cal P}_{tc}^{K\!K}}\right|_{\rm fact}=
\left[\frac{F_{B\pi}(M_K^2;0^+)}{F_{B K}(M_K^2;0^+)}\right]
\left[\frac{M_B^2-M_\pi^2}{M_B^2-M_K^2}\right]=0.79.
\end{equation} 
The numerical value in (\ref{SU(3)-fact2}) corresponds again to the 
light-cone sum-rule analysis performed in \cite{Ball} (with $\delta_{a1}=0$).
If we now use $\mbox{BR}(B^\pm\to\pi^\pm K)=(21.8\pm1.4)\times10^{-6}$ 
\cite{HFAG}, $\tau_{B^+}/\tau_{B_d}=1.086\pm0.017$, as well as 
(\ref{deltaR}) and (\ref{SU(3)-fact2}), (\ref{B-det2}) allows us to 
convert (\ref{B-min}) into the following lower bound:
\begin{equation}\label{BRmin2}
\mbox{BR}(B_d\to K^0\bar K^0)_{\rm min} = \left(1.36^{+0.18}_{-0.21}\right)   
\times\left[\frac{F_{B K}(M_K^2;0^+)}{0.331}\frac{0.258}{F_{B\pi}
(M_K^2;0^+)}\right]^2\times 10^{-6}.
\end{equation}
In comparison with (\ref{B-extr-1}), the advantage of (\ref{B-det2}) is 
obviously that the $B\to\pi\pi$ analysis enters only through $\delta R$, 
which has a small numerical impact. This feature is nicely reflected by 
the errors of (\ref{BRmin2}), which are considerably reduced with respect
of (\ref{BRmin1}), while the central values are very similar. On the 
other hand, we have to rely on the neglect of the penguin annihilation 
topologies in $B^0_d\to K^0\bar K^0$, so that (\ref{B-extr-1}) is 
conceptually more favourable. 

In view of the different assumptions entering (\ref{BRmin1}) and 
(\ref{BRmin2}), we consider it as very remarkable to arrive at such a 
consistent picture (see also (\ref{xi-SU3-det})). Looking at (\ref{BR-CP-av}), 
we observe that these {\it lower} SM bounds are very close to the current 
experimental {\it upper} bound, thereby suggesting that the observation of 
the decay $B_d^0\to K^0\bar K^0$ at the $B$ factories is just ahead of us. 
If we assume again that the penguin annihilation contributions to 
$B^0_d\to K^0\bar K^0$ are small, the decay $B^+\to K^+\bar K^0$ has
a very similar branching ratio; the current experimental upper bound
is given by $2.5\times 10^{-6}$ ($90\%$ C.L.) \cite{HFAG}. The latter 
mode is the $U$-spin counterpart of $B^+\to \pi^+K^0$, i.e.\ both
channels are related to each other by interchanging all down and
strange quarks, and was discussed in the context of dealing with the 
parameter $\rho_{\rm c}$ \cite{BFRS,rho-c-det}.

\boldmath
\subsection{Upper Bounds on $\langle B\rangle$ and 
$\rho_{K\!K}$}\label{ssec:B-rho-bounds}
\unboldmath
It is also interesting to convert the experimental upper bound in 
(\ref{BR-CP-av}) into upper bounds for $\langle B\rangle$. Using
(\ref{B-extr-1}) and (\ref{B-det2}), we obtain
\begin{equation}\label{B-bound1}
\langle B \rangle_{\rm max} = \left(0.88^{+0.90}_{-0.57}\right) \times
\left[\frac{F_{B\pi}(M_\pi^2;0^+)}{0.258}
\frac{0.331}{F_{B K}(M_K^2;0^+)}\right]^2 \times 
\left[{\mbox{BR}(B_d\to K^0\bar K^0)\over 1.5 \times 10^{-6}}\right]
\end{equation}
and
\begin{equation}\label{B-bound2}
\langle B \rangle_{\rm max} = \left(0.91^{+0.10}_{-0.09}\right) \times 
\left[\frac{F_{B\pi}(M_K^2;0^+)}{0.258}
\frac{0.331}{F_{B K}(M_K^2;0^+)}\right]^2 \times
\left[{\mbox{BR}(B_d\to K^0\bar K^0)\over 1.5 \times 10^{-6}}\right],
\end{equation}
respectively. We observe that the numerical values in (\ref{B-bound1})
and  (\ref{B-bound2}) are very close to the lower bound in (\ref{B-min}), 
which is of course no surprise because of the discussion given above. 
The interesting aspect of an upper bound for $\langle B \rangle$
is that it allows us to obtain an upper bound for $\rho_{K\!K}$ with
the help of the following relation:
\begin{equation}\label{rhoKK-bound}
\rho_{K\!K}< \left|\cos\gamma\right|+
\sqrt{\langle B \rangle_{\rm max}-\sin^2\gamma},
\end{equation}
where the central values in (\ref{B-bound1}) and (\ref{B-bound2}) 
correspond for $\gamma=65^\circ$ to $\rho_{K\!K}<0.66$ and 
$\rho_{K\!K}<0.72$, respectively, but the uncertainties remain 
sizeable. 

Looking at (\ref{rho-c}), we see that these upper bounds for 
$\rho_{K\!K}$ imply that $\rho_{\rm c}$ is actually tiny, in accordance 
with the discussion after (\ref{ACP-B+pi+K0}). In \cite{BFRS}, the 
experimental upper bound for BR$(B^\pm\to K^\pm K)$ discussed above was 
converted into $\rho_{\rm c}<0.1$ with the help of the $U$-spin relation 
to BR$(B^\pm\to \pi^\pm K)$, which would conversely correspond to 
$\rho_{K\!K}\lsim 2$. Consequently, (\ref{rhoKK-bound}) yields stronger 
constraints on this parameter.

\boldmath
\subsection{Comments on a Different Avenue: Extraction of $\gamma$}
\unboldmath
The analysis discussed above depends on the value of $\gamma$. This 
parameter enters explicitly in the corresponding formulae, but also 
implicitly through the values of $d$ and $\theta$ in (\ref{d-theta-extr}), 
which follow from the direct and mixing-induced CP asymmetries of 
$B_d\to\pi^+\pi^-$ and are actually functions of $\gamma$ \cite{BFRS}. 
However, if we do {\it not} assume that $\gamma$ is known, it is easy to 
see that the determination of the three $B_d\to K^0\bar K^0$ observables 
${\cal A}_{\rm CP}^{\rm dir}$, ${\cal A}_{\rm CP}^{\rm mix}$ and 
$\langle B \rangle$ allows us to extract simultaneously 
$\rho_{K\!K}$, $\theta_{K\!K}$ {\it and} $\gamma$, up to discrete 
ambiguities. This feature is not surprising, since it was suggested 
in \cite{BF-BdKK} to complement the CP-violating $B_d\to\pi^+\pi^-$ 
asymmetries with the observables provided by $B_d\to K^0\bar K^0$ to 
deal with the famous penguin problem in the former channel and to 
determine the angle $\alpha$ of the unitarity triangle. We have just 
encountered a different implementation of this strategy. Alternative 
methods to extract $\gamma$ from $B_d\to K^0\bar K^0$ were proposed 
in \cite{BKK-Uspin}, combining this channel with its $U$-spin partner 
$B_s\to K^0\bar K^0$.

\boldmath
\section{Correlations with the $B\to\pi\pi$ System}\label{sec:Bpipi-BKK}
\begin{figure}
\vspace*{0.3truecm}
\begin{center}
\includegraphics[width=11cm]{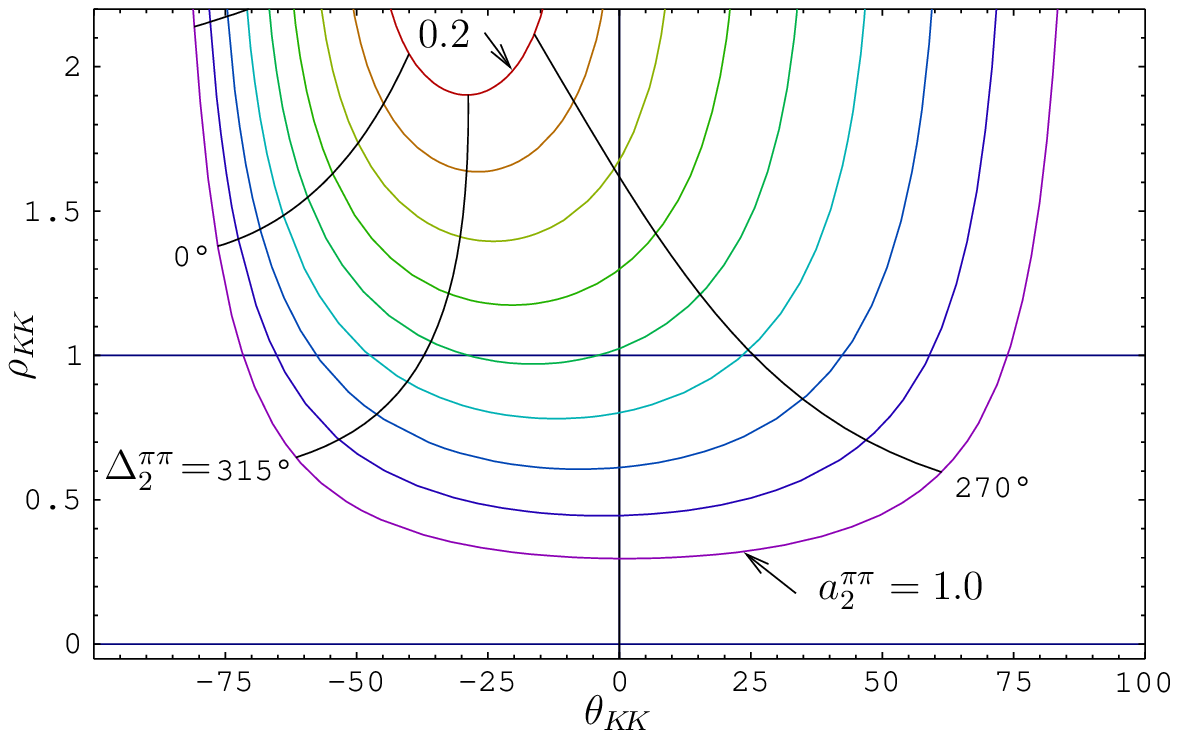}
\end{center}
\caption{The contours in the $\theta_{K\!K}$--$\rho_{K\!K}$ plane 
corresponding to the central values of $(d,\theta)$ and $(x,\Delta)$ 
in (\ref{d-theta-extr}) and (\ref{x-Delta-extr}), respectively, 
for various values of $a_2^{\pi\pi}$ and $\Delta_2^{\pi\pi}\in
[0^\circ,360^\circ]$.}
\label{fig:rhoKK-thetaKK}
\end{figure}

\unboldmath
The decay $B^0_d\to K^0\bar K^0$ will also allow us to obtain valuable 
insights into the substructure of the $B\to\pi\pi$ system. In the analysis 
of these decays in \cite{BFRS}, another hadronic parameter,
\begin{equation}
xe^{i\Delta}\equiv \frac{{\cal C}_{\pi\pi}+({\cal P}_{tu}^{\pi\pi}-
{\cal E}_{\pi\pi})}{{\cal T}_{\pi\pi}-({\cal P}_{tu}^{\pi\pi}-
{\cal E}_{\pi\pi})},
\end{equation}
was introduced, where ${\cal C}_{\pi\pi}$ and ${\cal T}_{\pi\pi}$ are 
the strong amplitudes of colour-suppressed and colour-allowed 
tree-diagram-like topologies, respectively, ${\cal P}_{tu}^{\pi\pi}\equiv
{\cal P}_t^{\pi\pi}-{\cal P}_u^{\pi\pi}$ is defined in analogy to 
${\cal P}_{tc}^{\pi\pi}$, and ${\cal E}_{\pi\pi}$ describes an
exchange topology. In analogy to the determination of $d$ and 
$\theta$ (see (\ref{d-theta-extr})), $x$ and $\Delta$ can also be 
extracted from the $B\to\pi\pi$ data, with the following
result:\footnote{There is also a second solution for
$(x,\Delta)$, which is, however, disfavoured by the $B\to\pi K$ data.}
\begin{equation}\label{x-Delta-extr}
x=1.22^{+0.26}_{-0.21},\quad \Delta=-\left(71^{+19}_{-26}\right)^\circ.
\end{equation}
If we now introduce the ``colour-suppression'' parameter
\begin{equation}
a_2^{\pi\pi}e^{i\Delta_2^{\pi\pi}}\equiv
\frac{{\cal C}_{\pi\pi}}{{\cal T}_{\pi\pi}},
\end{equation}
neglect the exchange amplitude ${\cal E}_{\pi\pi}$, which is expected
to play a minor r\^ole and can be explored with the help of the
$B_d\to K^+K^-$, $B_s\to\pi^+\pi^-$ system \cite{BFRS}, and use
the $SU(3)$ flavour symmetry of strong interactions, we obtain
\begin{equation}\label{rho-KK-a2}
\rho_{K\!K}e^{i\theta_{K\!K}}=\left[\frac{a_2^{\pi\pi}e^{i\Delta_2^{\pi\pi}}
-xe^{i\Delta}}{a_2^{\pi\pi}e^{i\Delta_2^{\pi\pi}}+1}\right]
\frac{e^{-i\theta}}{d}.
\end{equation}
In Fig.~\ref{fig:rhoKK-thetaKK}, we illustrate the resulting contours 
in the $\theta_{K\!K}$--$\rho_{K\!K}$ plane for various values of 
$a_2^{\pi\pi}$ and $\Delta_2^{\pi\pi}\in[0^\circ,360^\circ]$, taking also into 
account that values of $\rho_{K\!K}$ being significantly larger than 1 are 
disfavoured because of the discussion in Subsection~\ref{ssec:B-rho-bounds}.
In order to simplify the analysis, we have considered the central values 
of $(d,\theta)$ and $(x,\Delta)$ in (\ref{d-theta-extr}) and 
(\ref{x-Delta-extr}), respectively. By the time the CP-violating 
$B_d\to K^0\bar K^0$ observables can be measured, much more accurate 
determinations of these parameters will anyway be available. As soon 
as $\rho_{K\!K}$ and $\theta_{K\!K}$ are extracted from the 
$B_d\to K^0\bar K^0$ observables, (\ref{rho-KK-a2}) allows us to determine 
$a_2^{\pi\pi}$ and $\Delta_2^{\pi\pi}$ with the help of 
\begin{equation}
a_2^{\pi\pi}e^{i\Delta_2^{\pi\pi}}=\frac{xe^{i\Delta}+de^{i\theta}
\rho_{K\!K}e^{i\theta_{K\!K}}}{1-de^{i\theta}\rho_{K\!K}e^{i\theta_{K\!K}}}.
\end{equation}
Following \cite{BFRS}, we may then also determine the hadronic parameter
$\zeta_{\pi\pi} e^{i\Delta_\zeta^{\pi\pi}}\equiv 
{\cal P}_{tu}^{\pi\pi}/{\cal T}_{\pi\pi}$,
as well as ${\cal P}_{tc}^{\pi\pi}/{\cal T}_{\pi\pi}$, so that we are in 
a position to resolve the whole substructure of the $B\to\pi\pi$ system. 
In particular, we may then pin down the interference effects between 
the different hadronic penguin amplitudes, and may decide which one 
of the patterns suggested in the literature (see, for instance,
\cite{BFRS,Bpipi-charm}) is actually realized in nature.

\begin{figure}
\vspace*{0.3truecm}
\begin{center}
\includegraphics[width=11cm]{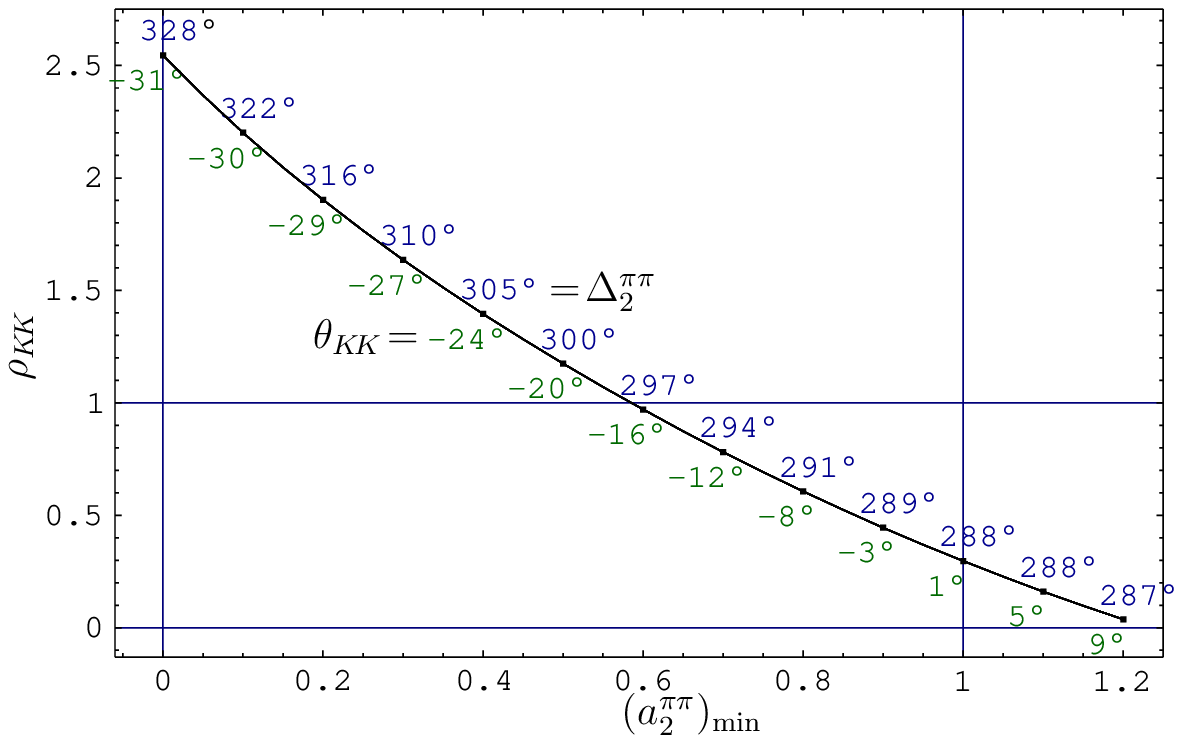}
\end{center}
\caption{The correlation between the upper bound for $\rho_{K\!K}$ and 
the corresponding lower bound for $a_2^{\pi\pi}$ with the associated 
values of $\theta_{K\!K}$ and $\Delta_2^{\pi\pi}$ for the case shown 
in Fig.~\ref{fig:rhoKK-thetaKK}.}\label{fig:rhoKKa2}
\end{figure}

If we look at Fig.~\ref{fig:rhoKK-thetaKK}, we observe that {\it upper} 
bounds for $\rho_{K\!K}$ correspond to {\it lower} bounds for $a_2^{\pi\pi}$,
as illustrated in Fig.~\ref{fig:rhoKKa2}. For $\rho_{K\!K}\lsim0.9$, we obtain 
$a_2^{\pi\pi}\gsim 0.6$. Consequently, the rather stringent upper bounds for 
$\rho_{K\!K}$ following from (\ref{rhoKK-bound}) require a sizeable 
deviation from the na\"\i ve value of 
$a_2^{\pi\pi} e^{i\Delta_2^{\pi\pi}}\sim 0.25$. This observation is
in accordance with discussion given in \cite{BFRS}, putting it on 
more solid ground. In this picture, we have {\it destructive} interference 
between the ${\cal P}_t^{\pi\pi}$ and ${\cal P}_c^{\pi\pi}$ amplitudes, 
whereas the interference between ${\cal P}_t^{\pi\pi}$ and 
${\cal P}_u^{\pi\pi}$ is {\it constructive}, with
$|{\cal P}_t^{\pi\pi}/{\cal T}_{\pi\pi}|\sim 
|{\cal P}_u^{\pi\pi}/{\cal T}_{\pi\pi}|\sim 0.25$. Moreover, 
$0.5\lsim a_2^{\pi\pi}\lsim 0.7$ with $\Delta_2^{\pi\pi}\sim290^\circ$
is suggested, where $\rho_{K\!K}$ is actually close to its current 
experimental upper bounds discussed in Subsection~\ref{ssec:B-rho-bounds}, 
as can be seen in Fig.~\ref{fig:rhoKKa2}. 

\begin{figure}
\vspace*{0.3truecm}
\begin{center}
\includegraphics[width=11cm]{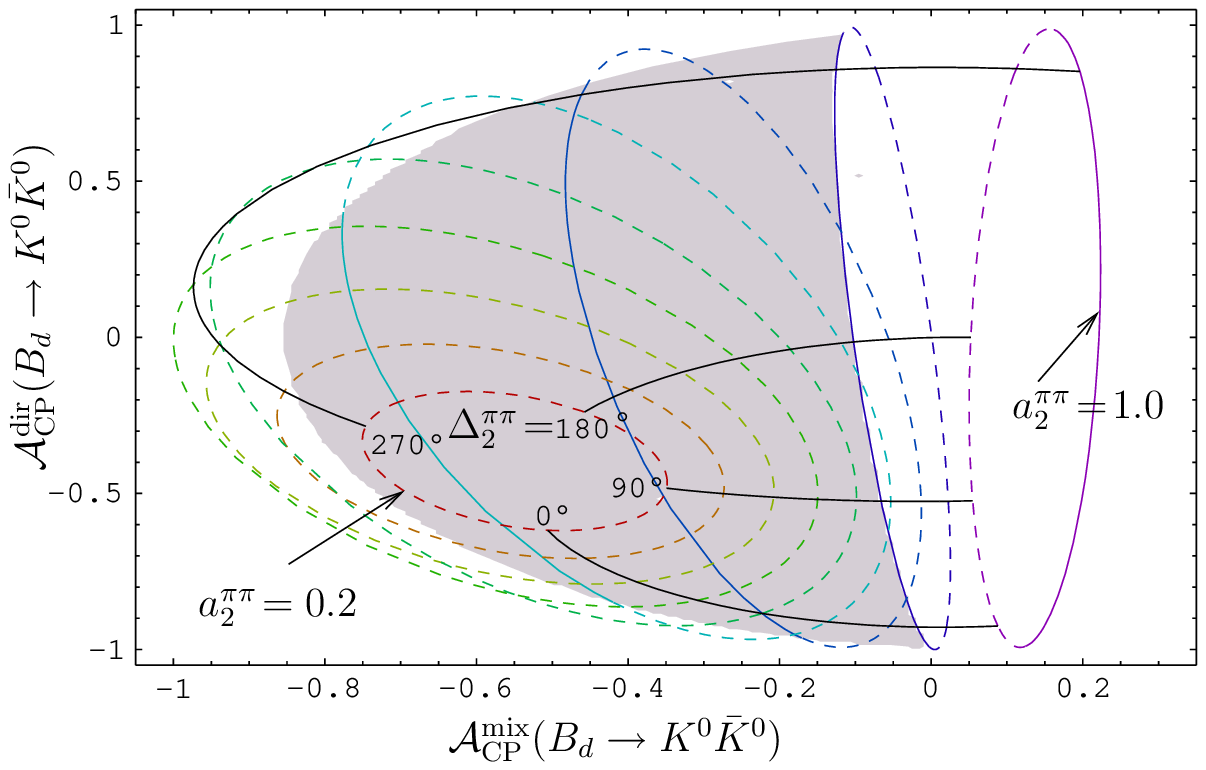}
\end{center}
\caption{The contours in the 
${\cal A}_{\rm CP}^{\rm mix}$--${\cal A}_{\rm CP}^{\rm dir}$ plane
corresponding to different values of $a_2^{\pi\pi}$ between 0.2 and 1.
The contours are drawn solid for $\rho_{K\!K}\le 0.9$ and dashed for
$\rho_{K\!K}> 0.9$.
The shaded region illustrates the area where  $a_2^{\pi\pi}<0.9$ and
$\rho_{K\!K}<0.9$.}\label{fig:Amix-Adir}
\end{figure}

\begin{figure}
\vspace*{0.3truecm}
\begin{center}
\includegraphics[width=11cm]{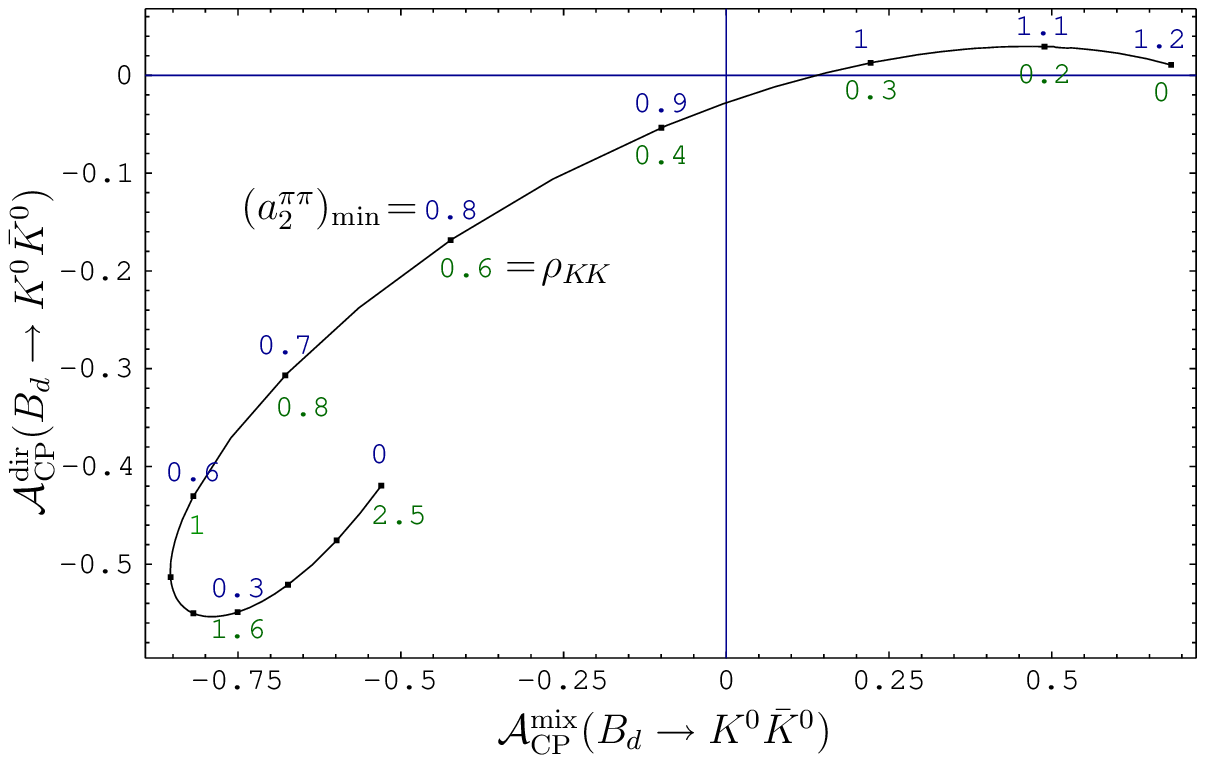}
\end{center}
\caption{The contour in the 
${\cal A}_{\rm CP}^{\rm mix}$--${\cal A}_{\rm CP}^{\rm dir}$ plane
arising for the limiting case illustrated in Fig.~\ref{fig:rhoKKa2}.
The numbers below and above the curve correspond to the given upper 
bound for $\rho_{K\!K}$ and the associated minimal value of 
$a_2^{\pi\pi}$, respectively.}\label{fig:Amix-Adir-rhomin}
\end{figure}

Let us finally come back to the CP-violating observables 
${\cal A}_{\rm CP}^{\rm dir}$ and ${\cal A}_{\rm CP}^{\rm mix}$
of the decay $B^0_d\to K^0\bar K^0$. In Fig.~\ref{fig:Amix-Adir}, 
we consider the ${\cal A}_{\rm CP}^{\rm mix}$--${\cal A}_{\rm CP}^{\rm dir}$
plane and show the contours for different values of $a_2^{\pi\pi}$, where
each point is parametrized by a given value of $\Delta_2^{\pi\pi}$. In 
accordance with our upper bounds for $\rho_{K\!K}$, we assume that 
$\rho_{K\!K}<0.9$; the contours are dashed where this bound is violated. 
The shaded region is calculated with the help of (\ref{rho-KK-a2}) for the 
central values of $(d,\theta)$ and $(x,\Delta)$ in (\ref{d-theta-extr}) and 
(\ref{x-Delta-extr}), respectively, imposing the constraints of 
$\rho_{K\!K}<0.9$ and $a_2^{\pi\pi}<0.9$. As far as the latter bound
is concerned, we allow for values being significantly larger than
the range discussed above to be on the conservative side. From the 
position of the contours it can be seen how this region changes for 
different upper bounds on $a_2^{\pi\pi}$. We observe that an interesting
pattern emerges, where {\it negative} values of the mixing-induced 
$B^0_d\to K^0\bar K^0$ CP asymmetry are preferred. In order to complement 
Fig.~\ref{fig:Amix-Adir}, we show in Fig.~\ref{fig:Amix-Adir-rhomin} the 
curve corresponding to the correlation between the lower bounds on 
$a_2^{\pi\pi}$ that are implied by upper bounds on $\rho_{K\!K}$, as 
illustrated in Fig.~\ref{fig:rhoKKa2}.

It should be noted that the analysis performed in this section -- and the 
pattern in the ${\cal A}_{\rm CP}^{\rm mix}$--${\cal A}_{\rm CP}^{\rm dir}$
plane -- do not depend on the $SU(3)$-breaking ratio of the $F_{B\pi}$ 
and $F_{B K}$ form factors that we encountered in Section~\ref{sec:char-SM}.
This quantity enters only implicitly when we impose the upper bounds
for $\rho_{K\!K}$ that are extracted from the current $B$-factory data.

\section{Conclusions}\label{sec:concl}
In our analysis of the penguin mode $B^0_d\to K^0\bar K^0$, we have
first shown that this channel can be efficiently characterized in the 
SM through a theoretically clean surface in the space of its observables 
${\cal A}_{\rm CP}^{\rm dir}$, ${\cal A}_{\rm CP}^{\rm mix}$ and 
$\langle B\rangle$. Whereas the CP asymmetries can straightforwardly be 
determined from time-dependent rate measurements, the extraction of 
$\langle B\rangle$ from the CP-averaged $B^0_d\to K^0\bar K^0$ branching 
ratio requires additional information. This can be obtained from the 
$B\to\pi\pi$ system with the help of the $SU(3)$ flavour symmetry, 
including the factorizable $SU(3)$-breaking corrections through an 
appropriate form-factor ratio; we have also discussed how 
insights into non-factorizable $SU(3)$-breaking corrections of the 
relevant hadronic penguin amplitudes can be obtained, and have shown 
that the current $B$-factory data are consistent with small effects, 
although the errors are still large. Alternatively, $\langle B\rangle$ 
can also be determined with the help of the CP-averaged $B^\pm\to\pi^\pm K$ 
branching ratio, requiring the additional assumption of small penguin 
annihilation contributions to $B^0_d\to K^0\bar K^0$. For our numerical
analysis, we have used the $SU(3)$-breaking form factor ratio obtained 
in a recent light-cone sum-rule calculation, which is consistent with
the BSW model; further analyses are desirable.

Following these lines, we pointed out that there is a lower bound for 
the CP-averaged $B_d\to K^0\bar K^0$ branching ratio within the SM, 
where the $B\to\pi\pi$ and $B^\pm\to\pi^\pm K$ avenues give remarkably 
consistent pictures. The interesting feature of this lower bound is that 
it is found to be very close to the current experimental upper bound. 
Consequently, we expect that the decay $B^0_d\to K^0\bar K^0$ will soon 
be observed at the $B$ factories. 

Finally, we have explored the interplay between $B^0_d\to K^0\bar K^0$
and the $B\to\pi\pi$ system, where the former channel allows us to
resolve the whole hadronic substructure of the latter modes. In 
particular, we have shown that upper bounds for $\rho_{K\!K}$ imply 
lower bounds for the colour-suppression factor $a_2^{\pi\pi}$, 
pointing to a sizeable deviation from the na\"\i ve value of 
$a_2^{\pi\pi} e^{i\Delta_2^{\pi\pi}}\sim 0.25$. Moreover, we have
analysed the impact on the allowed region in the plane of the
CP-violating $B_d\to K^0\bar K^0$ observables, and found that the 
current $B$-factory data have a preference for negative values of 
the corresponding mixing-induced CP asymmetry ${\cal A}_{\rm CP}^{\rm mix}$. 
By the time these quantities can be measured, we will have a much better 
picture of the parameters entering this analysis, allowing us to perform 
an interesting test of the SM description of $\bar b\to \bar d s\bar s$ FCNC 
processes, which are currently essentially unexplored. The full
implementation of these strategies should provide an interesting 
playground for an $e^+e^-$ super $B$-factory.

\vspace*{0.6truecm}

\noindent
{\bf Acknowledgements}\\
\noindent
The work presented here was supported in part by the German Bundesministerium 
f\"ur Bildung und Forschung under the contract 05HT4WOA/3 and the DFG project 
Bu.\ 706/1-2.

\newpage

\end{document}